\documentclass[floatfix,reprint,amsmath,amssymb,prapplied,superscriptaddress,aps]{revtex4-2}

\usepackage[dvipsnames]{xcolor}
\usepackage{graphicx}
\usepackage{amsmath,amssymb}
\usepackage{dcolumn}
\usepackage[utf8]{inputenc}
\usepackage[T1]{fontenc}
\usepackage{bm}
\usepackage{fontawesome5}
\usepackage{siunitx}
\usepackage{hyperref}
\usepackage{xcolor}
\usepackage{booktabs,longtable}
\usepackage{soul}
\colorlet{eli}{orange}

\begin{document}

\title{Engineering micro-disorder for macro-performance in magnetic nanoparticles}

\author{Jonathan Leliaert\textsuperscript{$\dagger$}}
\affiliation{DyNaMat, Department of Solid State Sciences, Ghent University, 9000 Ghent, Belgium}
\author{Elizabeth M. Jefremovas\textsuperscript{$\dagger$}}
\email{Corresponding author: elizabeth.jefremovas@uni.lu. \newline{$\dagger$}Both authors contributed equally to this work. }
\affiliation{Department of Physics and Materials Science, University of Luxembourg, 162A Avenue de la Faiencerie, L-1511 Luxembourg, Grand Duchy of Luxembourg}
\affiliation{Institute for Advanced Studies, University of Luxembourg, Campus Belval, L-4365 Esch-sur-Alzette, Luxembourg}

\date{\today}


\begin{abstract}
Spin disorder, inherent to magnetic nanoparticles, has traditionally been regarded as a detrimental feature, with materials-engineering efforts largely focused on producing ``perfect particles'' containing as few defects as possible. Alongside this pursuit of perfection, however, an alternative framework has emerged in recent years that reframes intra-particle disorder as an ``ugly duckling'' whose functional potential remains to be unlocked. In this Perspective, we review the emerging concept of disorder engineering in magnetic nanoparticles, identify its current challenges, and outline promising future directions.
From a theoretical standpoint, progress requires moving beyond the widely used macrospin approximation, which severely restricts the description of intra-particle degrees of freedom. Micromagnetic modelling, in contrast, treats magnetisation as a continuous vector field and thereby enables (i) the explicit representation of intra-particle degrees of freedom, linking microstructural features to internal magnetisation textures, and (ii) direct correspondence with polarized small-angle neutron scattering, an experimental technique that provides quantitative access to ensemble-averaged magnetic correlations on nanometre length scales. The field must now advance towards falsifiable and uncertainty-aware models with structurally motivated parameters and predictions that can be tested against independent experimental observables. The overarching goal is to establish quantitative relationships between particle structure, intra-particle magnetisation textures, and macroscopic functionality, thereby transforming spin disorder from an elusive hidden variable into an engineerable design parameter.
\end{abstract}

\maketitle


\section{Introduction}
In recent years, magnetic disorder in nanomagnetism has evolved from an unavoidable material imperfection into a design variable that can be exploited to enhance performance~\cite{lak2021embracing, mazza2024embracing, simonov2020designing}. This conceptual shift spans markedly different magnetic systems. In single-crystalline and polycrystalline materials, disorder has been shown to reshape Griffiths-like phases and collective magnetic behaviour~\cite{ghorai2022effect, rama2004site, marcano2025evolution}. In spin liquids, the interplay between disorder and frustrated interactions can stabilize unconventional quantum states relevant to quantum technologies~\cite{savary2017disorder, furukawa2015quantum, hu2021freezing, de2023momentum}. In confined magnetic films and nanostructures, frustration and disorder can modify the stability and dynamics of topological textures, including skyrmions and hopfions, with implications for unconventional and neuromorphic computing~\cite{raphael_confinement, jefremovas2025role, jefremovas2024experimental, kent2021creation, tran2025controlling, brems2025realizing, tanaka2026size}. Magnetic nanoparticles (MNPs), however, remain a notable exception. Despite their central role in nanomagnetism and magnetic nanotechnology, the controlled use of spin disorder to optimize MNP functionality remains an under-exploited playground. \newline

Disorder-dependent macroscopic behaviour has nevertheless been reported in a wide range of MNP systems, including iron oxide nanoflowers and cuboidal chains~\cite{jefremovas2026coercivity, jefremovas2025micromagnetic}, cobalt ferrite nanocubes and nanospheres~\cite{zakutna2022multiscale, zakutna2020field}, and spherical CoCr$_2$O$_4$ nanoparticles~\cite{zakutna2018noncollinear}. The central challenge lies therefore in establishing how a given nanoparticle microstructure determines its macroscopic functionality. Addressing this question requires resolving the intra-particle magnetic textures and identifying how they emerge from structural and magnetic disorder. Building the link between structural features, micromagnetic textures, and macroscopic performance therefore requires moving beyond structure-averaged descriptors. In particular, it is necessary to determine how crystallographic disorder, magnetic anisotropy, exchange coupling, and magnetostatic interactions collectively shape the internal magnetisation of individual particles. \newline

Iron oxide nanoflowers (NFs) provide a particularly suitable platform for addressing these questions. NFs are hierarchical particles assembled from exchange-coupled primary crystallites, typically 5--30~nm in size. Their internal complex architecture is a natural source of grain boundaries, distribution of crystallographic orientations, and structural defects, which together generate a spatially heterogeneous magnetic landscape. Advances in materials design have enabled a high degree of control over these intrinsically complex morphologies. As such, NFs can be synthesized through organic~\cite{vita2016tuning, ali2019novel, salas2012controlled}, aqueous~\cite{sugimoto1980formation, vereda2013control, lin2013understanding}, and polyol-based routes~\cite{gavilan2017colloidal, gavilan2021magnetic}. In particular, polyol synthesis offers substantial control over particle size, morphology, crystallinity, and defect density, while microwave-assisted approaches have enabled fast, greener, and scalable production of gram-scale quantities within hours~\cite{simeonidis2024toward}. The key opportunity is therefore not simply that NFs are structurally disordered, but that this disorder may be synthetically controlled and deliberately engineered.\newline

Despite (or perhaps because of) their defect-rich nanoarchitecture, iron oxide NFs display outstanding performance in magnetic induction heating, with applications spanning magnetic hyperthermia, nanocatalysis, and environmental remediation~\cite{gavilan2025magnetic, gallo2022unravelling, gavilan2021magnetic, gallo2024magnetic}. Increasing evidence indicates that this functionality cannot be understood solely in terms of particle size, composition, or saturation magnetisation, but is intimately connected to intra-particle magnetic degrees of freedom and disorder-mediated spin textures~\cite{bender2018relating, jefremovas2026coercivity}. However, developing a mechanistic understanding of how magnetic disorder governs macroscopic performance, while experimentally resolving the underlying intra-particle magnetic textures, remains a major open challenge. \newline

In this Perspective, we argue that progress requires a transition from effective, spatially averaged descriptions of MNPs towards texture-resolved and physically grounded multiscale models. Recent advances in GPU-accelerated micromagnetic tools, such as mumax3\cite{vansteenkiste2014design} and mumax+\cite{moreels2026mumax+} now make it possible to model experimentally relevant nanoparticles and directly connect structural complexity to internal spin textures. We propose iron oxide NFs as a model platform to examine how structural complexity can be translated into internal magnetic textures and, ultimately, into a functional response. Particular emphasis is placed on the complementary roles of micromagnetic simulations and polarized small-angle neutron scattering (SANS): micromagnetics provides access to real-space magnetisation configurations and allows competing microscopic hypotheses to be tested, whereas polarized SANS offers ensemble-averaged experimental access to magnetic correlations at mesoscopic resolution. \newline

Looking ahead, computational frameworks such as Magnum.np\cite{Bruckner2023MagnumNP} and Neuralmag\cite{Abert2025NeuralMag} raise the prospect of addressing inverse problems, where experimental magnetic measurements are used to infer the underlying spin textures and (spatial distribution of) material parameters. We propose that the combined use of micromagnetic simulations and SANS provides a particularly powerful route to establish quantitative structure-texture-property relationships and, ultimately, to transform intraparticle magnetic disorder from an unresolved source of variability into an actionable materials-design parameter.\newline

\section{Intra-particle magnetisation: Beyond the macrospin framework}

One of the main obstacles to harnessing engineered disorder in MNPs stems from the conceptual framework through which these particles are commonly described.  The prevailing theoretical framework describes MNPs under the macrospin approximation~\cite{nowak2005spin}, which suppresses intra-particle degrees of freedom and usually restricts magnetic disorder to undesired spin canting. This approximation provides a relatively accurate description for small particles \textit{i.e.} below the single domain limit, where the Stoner-Wohlfarth model~\cite{tannous2008stoner} successfully reproduces the macroscopic behaviour of the MNPs~\cite{jefremovas2021nanoflowers, marcano2022magnetic, mostarac2025thermal}. However, above the single domain limit (typically, around 50 nm for maghemite-rich particles), intra-particle disorder becomes significant, adding a layer of complexity that remains uncaptured within the boundaries of the macrospin approximation. We invite the reader to visualize this situation as if they would be asked to describe a summer sunset only by stating its average colour. The overall colour may be correct, yet it is the interplay among shadows, colour gradients and contrasts, that makes it unique.  \newline 

Micromagnetism expands this colour palette by replacing the single, macrospin magnetic moment with a spatially resolved magnetisation field. Within this framework, the MNP is no longer treated as a homogeneous magnetic block, but is discretized into computational cells whose dimensions must remain below the relevant micromagnetic length scales (notably, the exchange length) to preserve the mathematical validity of the continuum model. Typical cell sizes are few nm. The magnetisation is assumed to be practically uniform within each cell, while neighbouring cells interact through exchange and respond to local anisotropy and external magnetic fields. In addition, all cells interact with each other through the long-range magnetostatic interaction. It is this competition between short-range and long-range interactions that provides the richness in magnetisation textures in particles larger than the single-domain limit. Groups of cells can be assigned to distinct material regions, making it possible to represent the grain structure, spatially varying anisotropy, and heterogeneous magnetic parameters of polycrystalline nanoparticles~\cite{gavilan2017colloidal, mekseriwattana2025shape}. In this way, the nucleation and stabilization conditions of non-uniform magnetic states, including domain walls or vortices, can be evaluated based on specific material parameter combinations, rather than pre-assumed~\cite{jefremovas2026coercivity}.\newline

Atomistic approaches, including Monte Carlo and atomistic spin-dynamics simulations, can explicitly account for atomic-scale defects, surface coordination, and local exchange interactions~\cite{bender2018dipolar, lappas2019vacancy, adams2022magnetic, moya2024unveiling}. Their computational cost, however, generally restricts the accessible particle dimensions, parameter space and timescales, particularly when statistically representative cases must be considered. Micromagnetism therefore occupies a crucial intermediate level: it retains the internal magnetic degrees of freedom that are absent from macrospin models while remaining applicable to MNP on application-relevant scales.\newline

\begin{figure*}
\centering
\includegraphics[width=\linewidth]{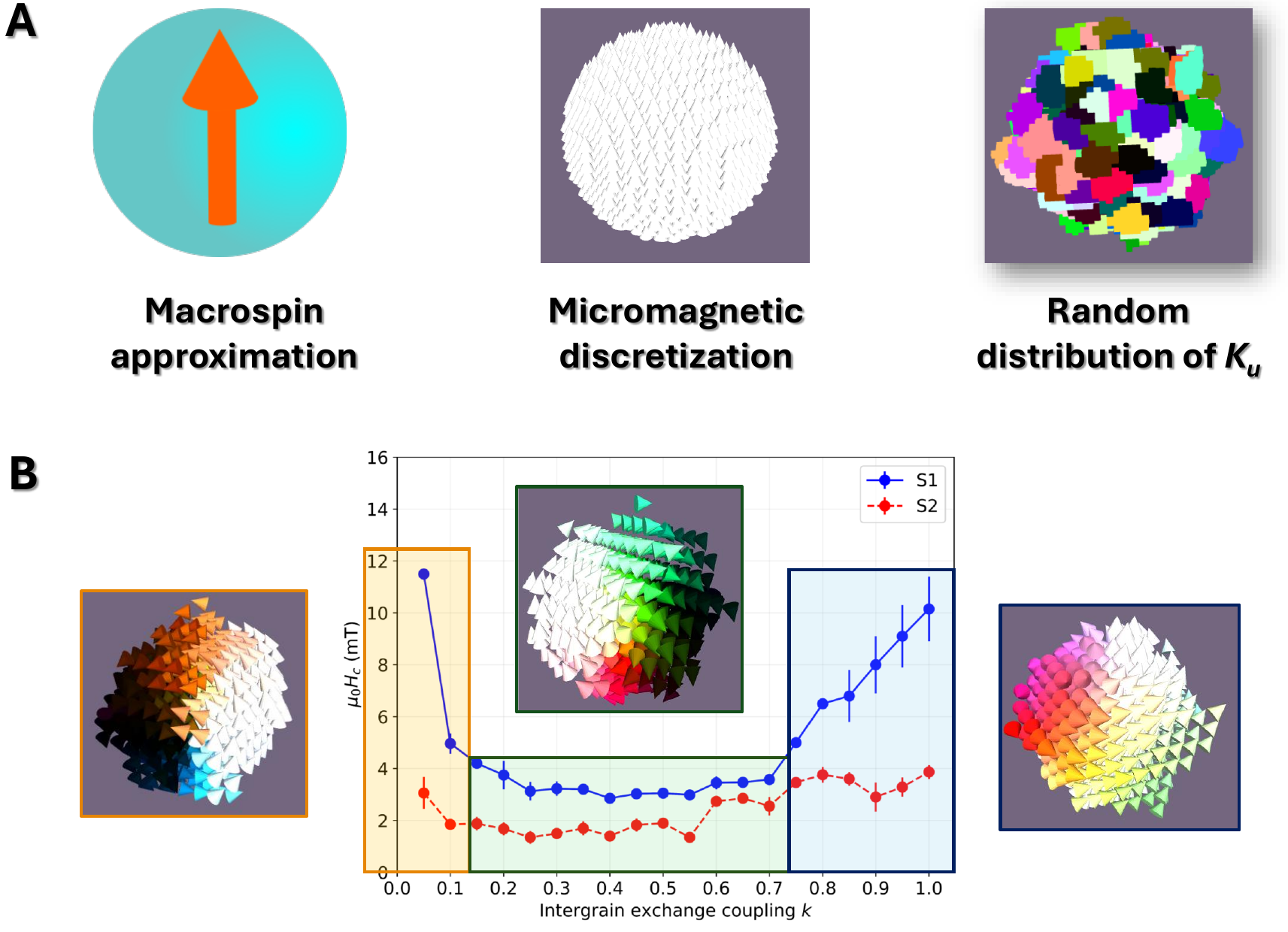}
\caption{\textbf{(A)} While the macrospin model reduces the particle magnetisation to a single effective magnetic moment, micromagnetics discretizes the particle into computational cells and resolves the spatially varying magnetisation. Mumax3 also allows to reproduce irregular morphologies and polycrystalline architectures observed experimentally~\cite{moya2024unveiling}. We have applied a Voronoi tessellation to define regions wherein the uniaxial anisotropy, $\vec{K_{u}}$, is defined along different directions, yielding a random distribution of $\vec{K_{u}}$ directions (represented in colors) within the NF. \textbf{(B)} Coercivity, $\mu_{0}H_{\mathrm{C}}$, as a function of the normalized intergrain exchange coupling, $k$, for a NF with nominal diameter $d=100~\mathrm{nm}$. All grains share the same uniaxial easy-axis direction, allowing the influence of intergrain exchange coupling to be examinated. Symbols represent averages over 20 structural realizations, with error bars indicating the dispersion among realizations. The results are separated according to the magnetization dynamics: the S1 states are vortex-core dominated, whereas S2 states are flux-closure-dominated. Representative remanent states at $k=0.05$, $0.45$, and $0.95$ illustrate the weak-, intermediate-, and strong-coupling regimes, respectively.}
\label{supplemental_ku_0}
\end{figure*}

Figure~\ref{supplemental_ku_0}\textbf{(A)} illustrates how micromagnetism can incorporate the polycrystalline intraparticle architecture of iron oxide NFs. In a conventional macrospin description, the internal magnetisation is reduced to a single effective magnetic moment, thereby excluding spatial variations in the magnetisation and the emergence of non-uniform spin textures. Micromagnetics instead discretizes the particle into computational cells and determines the equilibrium magnetisation by minimizing the total magnetic energy, including exchange, magnetostatic, anisotropy, and Zeeman contributions. Using MuMax3~\cite{Vansteenkiste2014}, a Voronoi tessellation can be employed to group these cells into discrete regions that represent the constituent nanocrystallites~\cite{Lel2014}. The resulting model can reproduce both the irregular external morphology and the internal polycrystalline architecture characteristic of NFs~\cite{moya2024unveiling, gavilan2017colloidal, gavilan2021size, gallo2022unravelling, gallo2024magnetic}. Individual nanocrystallites, hereafter referred to as grains, are represented as distinct regions and assigned local uniaxial easy-axis directions, $\vec{K_{u}}$. This grain-resolved description translates structural heterogeneity into a spatially varying anisotropy landscape and enables its consequences for the intraparticle magnetisation to be evaluated. \newline 

Besides particle's morphology and internal polycrystallinity, an additional source of intraparticle magnetic disorder originates from the magnetic coupling across grain boundaries. To represent this contribution, a dimensionless intergrain exchange factor, $k$, is introduced, such that the exchange stiffness across a grain boundary is written as $kA$, with $0\leq k\leq 1$~\cite{jefremovas2026coercivity, moya2024unveiling}. The limiting values $k=0$ and $k=1$ correspond, respectively, to fully decoupled grains and to boundaries across which the bulk exchange stiffness is preserved. Although $k$ constitutes an effective rather than directly measurable microscopic parameter, it provides a means of probing the grain-boundary-mediated pinning landscape governing collective reversal and remanent magnetic textures. Importantly, this landscape is indirectly experimentally accessible through synthesis. Control over the characteristic grain size has already been demonstrated~\cite{gavilan2017colloidal}, thereby allowing the number and density of intraparticle grain boundaries to be tailored. Thus, although synthesis does not directly tune $k$, it can modify the occurrence of the pinning sites whose magnetic effect is effectively captured by this parameter.\newline

Recently, we performed a micromagnetic study examinating the coercivity as a function of the NF size~\cite{jefremovas2026coercivity}, revealing a secondary maximum in the coercivity of NFs. This feature is interpreted as arising from the competition between two distinct contributions to the energy landscape: the anisotropy generated by the distribution of local easy-axis directions, and the pinning associated with variations in intergrain exchange coupling. Their example exposes a broader challenge for disorder-resolved nanomagnetism: different microscopic forms of disorder can produce similar changes in a macroscopic hysteresis loop. It remains, however, challenging to discriminate the role of each of them. While morphology, crystallinity, and crystallographic orientation can be constrained experimentally using X-ray diffraction and high-resolution transmission electron microscopy~\cite{jefremovas2021nanoflowers, moya2024unveiling}, the effective exchange coupling across individual grain boundaries remains considerably less accessible.\newline

Nonetheless, we propose to use coercivity as a useful guiding observable. Once particle morphology, grain size, and anisotropy distributions are independently constrained, comparison between experimental hysteresis loops and micromagnetic calculations can narrow the range of exchange parameters compatible with the measurements. Such an approach should not be regarded as a unique inversion of coercivity into an exchange constant: coercivity is influenced simultaneously by anisotropy, morphology, thermal activation, magnetostatic interactions, and the reversal pathway. Rather, we envisage it as part of an inference strategy in which macroscopic magnetic measurements constrain the model parameter space, while texture-sensitive techniques such as polarized SANS provide additional information capable of discriminating between otherwise degenerate internal configurations. \newline

Figure~\ref{supplemental_ku_0}\textbf{(B)} provides an illustrative example of how this parameter may be constrained. The calculations consider a NF with a nominal diameter of $d=100~\mathrm{nm}$ and average the coercivity over 20 structural realizations. To isolate the contribution of intergrain exchange from that of anisotropy disorder, all grains are assigned a common easy-axis direction, $\vec{K_{u}}=(0.0872,0,0.9962)$ close to parallel with the applied field $\vec{H}_{z}$, while $k$ is varied from fully decoupled grains to bulk-like exchange coupling. The resulting remanent configurations separate into two broad families, depending on the magnetization reversal pathway. For the case of S1 realizations, the vortex-core occupies more than 1/3 of the total particle volume, and the magnetization reversal is driven purely along the field axis ($z$ direction). For the case of S2 realizations, the flux-closure part takes over 1/3 of the total magnetic moments, and the magnetization reversal occupies intermediate positions perpendicular to the field axis~\cite{jefremovas2026coercivity}. The latter consequently tend to exhibit lower coercivity. This distinction is already instructive from a modelling perspective: particles with nominally identical morphology and material parameters can follow different reversal pathways depending on the detailed realization of their internal structure. Ensemble-averaged magnetic observables therefore encode not only parameter values, but also the statistical distribution of accessible magnetic textures. The interpretation of intra-particle textures based solely on macroscopic metrics yields potentially misleading interpretations, calling for the need of microscopically-resolved experimental techniques. \newline

Within the S1 family, the calculations suggest three exchange-coupling regimes. In the weak-coupling limit, $k\lesssim0.05$, the grains reverse largely independently and the coercivity remains comparatively high, reaching values of approximately $\mu_{0}H_{\mathrm{C}}\sim12~\mathrm{mT}$. At intermediate coupling, approximately $0.05<k<0.75$, exchange promotes more collective reversal and produces a broad region of reduced coercivity. In the strong-coupling regime, $k\gtrsim0.75$, the grain boundaries do not provide sufficiently strong pinning sites, and the NF behaves increasingly as an internally-continuous magnetic object. Consequently, the vortex core extends over a larger fraction of the magnetic volume and carries a larger uncompensated moment, yielding an increase in coercivity. These regimes provide a physically plausible interpretation of the non-monotonic coercivity, rather than a unique reconstruction of the underlying reversal mechanism. Their broader value lies in demonstrating that exchange disorder need not merely broaden or suppress the magnetic response: by reorganizing the internal texture and its reversal pathway, it can generate qualitatively new macroscopic behaviour. This is precisely the type of mechanism that remains inaccessible to a macrospin description. \newline

The example also highlights both the power and the present limitations of micromagnetic inference. Simulations provide a framework in which structural hypotheses can be translated into experimentally testable magnetic consequences, but their predictive value depends on independent experimental constraints. Existing studies reporting coercivity in iron oxide NFs predominantly concern polycrystalline structures~\cite{rai2026polarized, moya2024unveiling, jefremovas2021nanoflowers, gavilán2017formation, storozhuk2021stable}, making it difficult to isolate the respective contributions of morphology, crystallographic misorientation, grain-boundary density, and exchange coupling. A particularly valuable direction for future synthesis would therefore be the preparation of matched particle series in which one structural variable is modified at a time. These could include particles with comparable diameter and external morphology but systematically varied grain size, crystallographic alignment, defect density, or grain-boundary chemistry, together with single-crystalline particles of similar dimensions as reference systems. Such samples would provide substantially stronger benchmarks for micromagnetic models than comparisons between particles that differ simultaneously in size, shape, and crystallinity. More importantly, they would help determine whether an effective parameter such as $k$ corresponds to a reproducible materials property or merely compensates for unresolved structural complexity. \newline

In our view, the applicability of the macrospin approximation should not only be determined by a universal particle-size threshold alone, but by whether exchange interactions are sufficiently strong to maintain an approximately uniform magnetisation throughout the magnetic volume. Even for ``single-domain'' particles, both Z\'akutn\'a et al. and Jefremovas et al. have shown the penetration of magnetic order into disordered regions, yielding the modification of the macroscopic particle response~\cite{zakutna2020field, jefremovas2025micromagnetic}. \newline

Micromagnetism therefore provides the natural framework for describing MNPs beyond the uniformly magnetized regime. Emerging numerical platforms, including MuMax+~\cite{moreels2026mumax+}, are extending the range of accessible material descriptions towards increasingly complex magnetic systems, including antiferromagnetic order, magnetoelastic coupling, and non-collinear configurations. \newline

\paragraph{GPU performance in micromagnetic simulations}
Having established micromagnetic simulations as our suggested tool of choice to investigate the magnetisation dynamics of non-uniformly magnetized nanoparticles with complex geometries and spatially varying material parameters, including disorder, it is worth revisiting the discussion on its performance presented in the 2019 perspective \textit{Tomorrow’s micromagnetic simulations}\cite{leliaert2019tomorrow}. Since then, the mumax3 benchmark dataset has expanded substantially, as shown in Fig. \ref{fig:fast_computers}. Details of the full dataset are available in the appendix.\\
 
In 2019, we estimated an average performance increase of approximately 30\%/year, corresponding to a doubling time of 2.6 years. 
The benchmark results continue to show an approximately exponential increase in throughput, although the rate depends on the GPU class. 
The updated fits show that the top-end desktop GPUs doubled in performance every 2.9 years, whereas the high performance computing (HPC) datacenter GPUs showed a doubling time of only 2.2 years, consistent with their increasingly higher memory bandwidth, which remains a major performance constraint for large micromagnetic simulations. Mobile GPUs show a less systematic trend and remain generally slower than their desktop counterparts, reflecting their tighter power and thermal constraints.

Overall, these results point to a Moore's-law-like increase in micromagnetic throughput. The practical consequence is not only shorter runtimes, but also access to larger and more realistic simulations. 
Three-dimensional nanostructures can be discretized more finely, enabling interfaces (e.g. between material grains), complex geometries (e.g. nanoflowers\cite{gavilan2017colloidal}, or even Rubik-like nanocubes cubes\cite{rizzo2026gram}), and spatial variations in magnetic parameters to be represented more explicitly and reducing the need to absorb them into effective material parameters.\newline

Increased throughput makes longer simulations, broad parameter sweeps, and ensemble calculations over disorder realizations practically feasible. For magnetic nanoparticles, this is crucial because thermal fluctuations\cite{leliaert2017adaptively}, structural disorder, and particle-to-particle variability often require many repeated simulations. The cost of any single simulation is therefore no longer the main limitation, and the cumulative cost of large statistical ensembles are becoming more important. \newline


\begin{figure*}
  \includegraphics[width=0.9\linewidth]{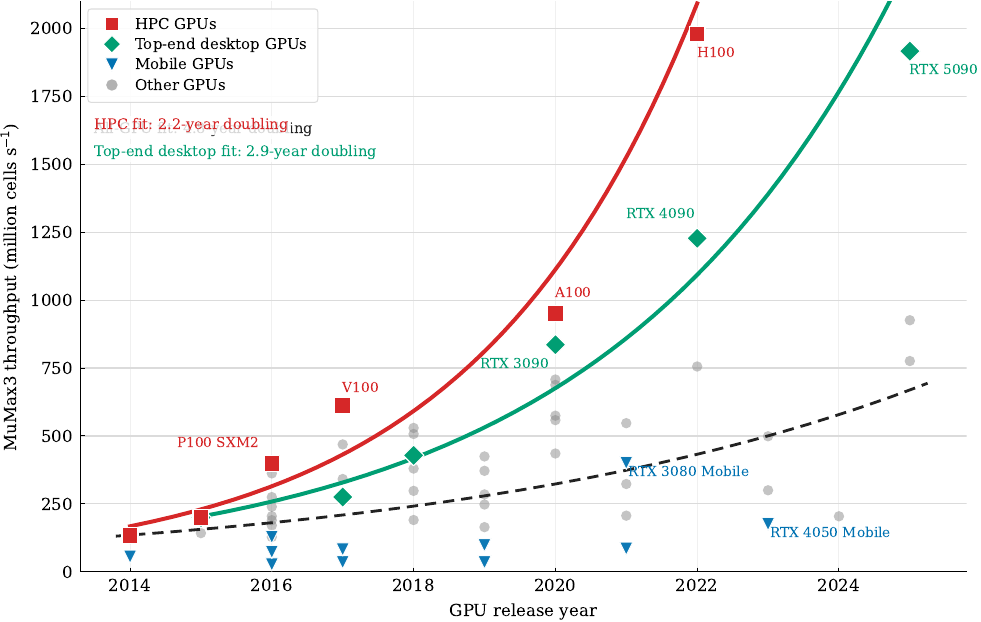}
  \caption{Evolution of GPU throughput for a $\sim$~4 million cell mumax3 benchmark. Red squares represent high performance computing (HPC), datacenter GPUs, green diamonds representative top-end desktop GPUs, blue triangles mobile GPUs, and grey circles show other benchmarked GPUs. Lines are least-squares fits in throughput, corresponding to empirical doubling times.}
  \label{fig:fast_computers}
\end{figure*}


\section{How to resolve intra-particle magnetisation experimentally}
Establishing physically grounded relationships between structural features and magnetic disorder requires access to the magnetisation at nanometre length scales comparable to the characteristic exchange and magnetic-correlation lengths. At these scales, spatial variations in the magnetisation emerge from the competition between exchange, anisotropy, magnetostatic, and Zeeman energies. 

Several microscopy techniques can provide such information. Magnetic force microscopy (MFM) and X-ray photoemission electron microscopy (XPEEM) can resolve magnetic textures with high spatial resolution, but their sensitivity is largely restricted to surface or near-surface regions: MFM measures the stray field above the sample, whereas the limited escape depth of photoelectrons makes XPEEM intrinsically surface sensitive~\cite{marcano2022magnetic, marques2024magnetic}. Transmission-based techniques, such as magnetic transmission X-ray microscopy and Lorentz transmission electron microscopy, provide access to the projected magnetisation of individual MNPs, but often under restrictive sample-preparation and environmental conditions~\cite{campanini2015lorentz, moya2024unveiling, marcano2022magnetic}. More recently, photothermal magnetic circular dichroism~\cite{spaeth2022imaging, adhikari2024single} and single-NV-center magnetometry have emerged as particularly powerful approaches for magnetic nanoparticles. PT-MCD probes magnetisation dynamics through differential absorption of circularly polarized light, while NV microscopy enables non-invasive imaging of static and dynamic magnetic textures with nanometre spatial resolution and sensitivity over a broad frequency range~\cite{everaert2025ac, finco2023single, rickhaus2024antiferromagnetic,celano2024probing, acosta2009diamonds, mathes2024nitrogen, richards2025time}\newline

Despite (or perhaps even due to) their excellent spatial resolution, these techniques predominantly provide information at the single-particle level. As a result, obtaining statistically meaningful measurements of magnetic disorder across large nanoparticle ensembles remains challenging, motivating the need for complementary ensemble-sensitive probes. Neutron-scattering techniques occupy a distinctive position within this methodological landscape. Because neutrons are electrically neutral, they penetrate deeply into matter, while their magnetic moment makes them directly sensitive to magnetic structures. Neutron scattering can therefore probe the internal magnetisation of bulk samples and nanoparticle ensembles with minimal perturbation of the magnetic state. It has provided quantitative access to nanoscale spin correlations~\cite{jefremovas2021observation, jefremovas2023magnetic} and to spatial variations of the magnetisation at intra-particle resolution~\cite{michels2021magnetic, zakutna2018noncollinear, zakutna2020field}. A major advantage is that these correlations are averaged over macroscopic sample volumes, typically corresponding to approximately 0.1~g or $10^{13}$ MNPs, thereby providing statistically representative metrics beyond the practical reach of real-space imaging techniques. In particular, small-angle neutron scattering (SANS) probes magnetic correlations over length scales extending approximately from a few nanometres to several hundred nanometres, while remaining compatible with applied magnetic fields, variable temperatures, and a variety of sample conditions and environments, including powder, single crystal or suspension~\cite{muhlbauer2019magnetic}. Major advances have been achieved both in the theoretical description of magnetic SANS~\cite{michels2021magnetic} and in its application to MNP ensembles~\cite{jefremovas2023magnetic, jefremovas2022magnetic, bender2018relating, bender2019supraferromagnetic, bender2018dipolar, zakutna2020field}. More recently, experimental and theoretical developments have extended the capabilities of SANS through polarization analysis (P-SANS), enabling a more selective decomposition of the magnetic scattering contributions~\cite{vivas2020toward, zakutna2020field, rai2026polarized}.\newline

Among these approaches, longitudinal polarization analysis is particularly valuable for studying weak spin-misalignment scattering. In this configuration, the incident and analysed neutron polarizations are defined with respect to the same guide-field direction, yielding four cross-sections conventionally denoted $++$, $--$, $+-$, and $-+$. The spin-flip channels, $+-$ and $-+$, are free from nuclear coherent scattering and selectively probe magnetic scattering associated with magnetisation components transverse to the neutron-polarization direction\footnote{We refer the reader to Ref.~\cite{michels2021magnetic} for extended discussions and derivations of the mathematical scaffold of polarized SANS cross-sections.}. While P-SANS involves substantially more demanding data acquisition and analysis than unpolarized SANS, the magnetic selectivity of its spin-flip channels enables quantitative access to weak transverse correlations and non-uniform intra-particle magnetisation textures that would otherwise be masked by dominant nuclear scattering contributions. \newline

\begin{figure*}
\centering
\includegraphics[width=\linewidth]{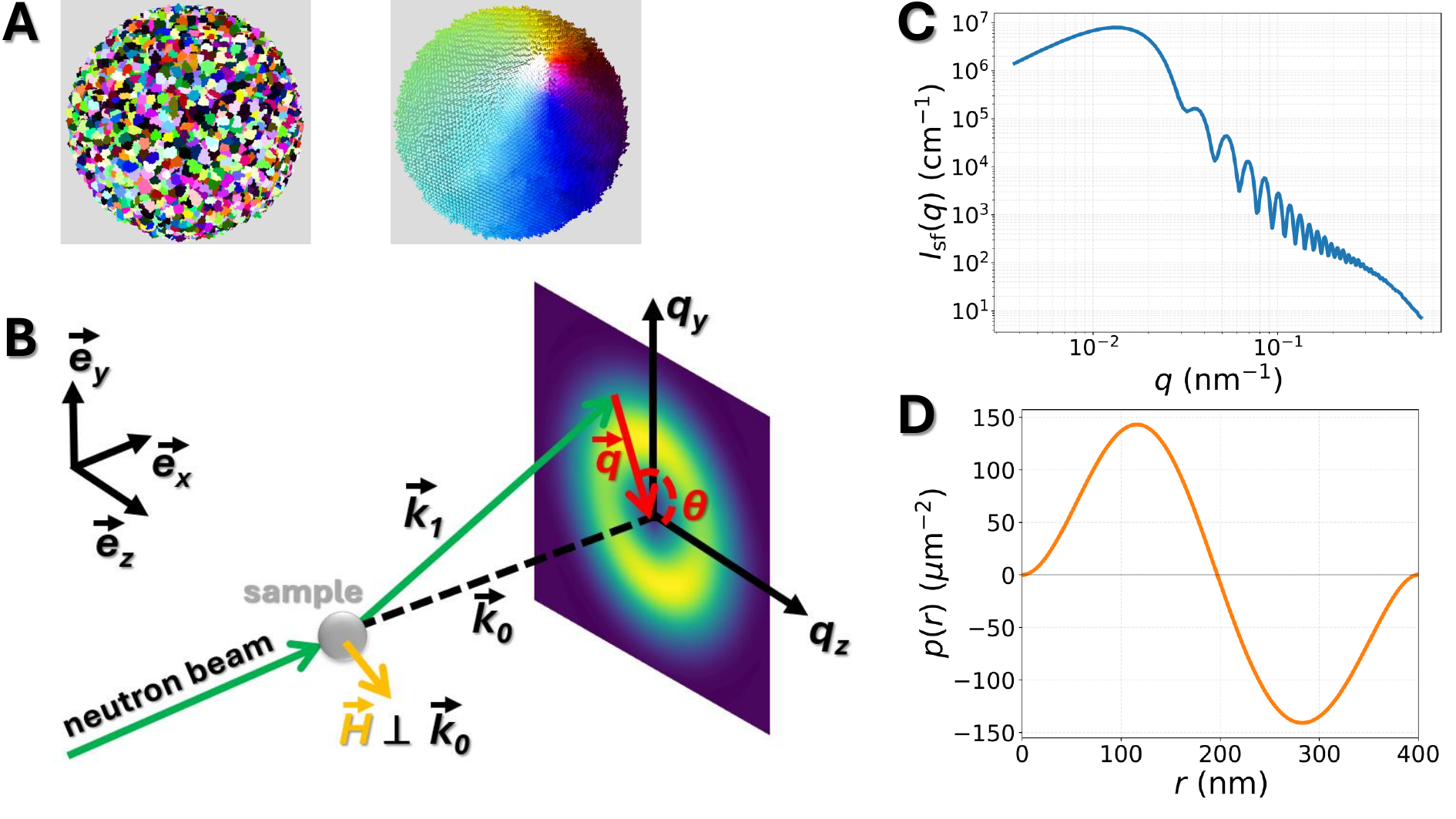}
\caption{\textbf{(A)} Multicore NF configuration used to calculate the scattering observables. It consists of a compact $400~\mathrm{nm}$ multicore aggregate assembled from $10~\mathrm{nm}$ single-domain nanocrystallites (uniaxial anisotropy directions represented by colours in the left-side snapshot), with a randomly distributed non-magnetic void fraction of 10\% and an intergrain exchange factor $k=0.25$. The remanent magnetisation (right-side) relaxes into a spherical vortex configuration. \textbf{(B)} Schematic representation of the transverse-field polarized-SANS geometry. The detector shows the simulated two-dimensional spin-flip SANS cross-section calculated from the micromagnetic magnetisation field. \textbf{(C)} Azimuthally averaged spin-flip scattering intensity, $I_{\mathrm{sf}}(q)$, displaying a characteristic suppression at low $q$. \textbf{(D)} Corresponding spin-flip pair-distance distribution function, $p(r)$, showing a sign change at approximately half the NF diameter. All scattering observables are numerically calculated using NuMagSANS software package~\cite{adamsjac2026}.}
\label{SANS}
\end{figure*}

How can P-SANS help to elucidate intra-particle spin disorder in multicore nanoarchitectures? We discuss herein an example by using a NF structure with a total diameter of approximately $400~\mathrm{nm}$ as a case example. The NF is assembled from single-domain nanocrystallites with a characteristic individual nanocrystallite size of $10~\mathrm{nm}$ (see Figure~\ref{SANS}\textbf{(A)}, left-side). A randomly distributed non-magnetic void fraction of 10\% is introduced to reproduce the internal empty regions commonly observed in polyol-synthesized aggregates~\cite{moya2024unveiling, gavilan2017colloidal, gavilan2021size, gallo2022unravelling, gallo2024magnetic}. Following recent studies indicating a reduction of the coupling at inter-grain boundaries~\cite{jefremovas2026coercivity, moya2024unveiling}, the intergrain exchange factor is fixed to $k=0.25$. As it can be observed from the micromagnetic real space representation shown in Figure~\ref{SANS}\textbf{(A)} (right-side), the internal magnetic texture relaxes into a vortex-like configuration at remanence.\newline

Figure~\ref{SANS}\textbf{(B)} illustrates a representative polarized-SANS geometry. The incident neutron beam, with wavevector $\mathbf{k}_{0}$, propagates along the $\mathbf{e}_{x}$ direction of the laboratory coordinate system. The scattered wavevector is denoted by $\mathbf{k}_{1}$, and the momentum-transfer vector is defined as $\mathbf{q}=\mathbf{k}_{1}-\mathbf{k}_{0}$. In the transverse-field geometry (most conventional case), the external magnetic field is applied perpendicular to the incident beam, $\mathbf{H}\parallel\mathbf{e}_{z}$. Modern SANS instruments can combine polarization analysis with broad reciprocal-space coverage and applied magnetic fields from few mT to few T~\cite{dewhurst2016small}. Using the state-of-the-art NuMagSANS software~\cite{adamsjac2026}, we have calculated the two-dimensional spin-flip SANS cross-section (ring-like pattern observed in the detector). The NuMagSANS software~\cite{adamsjac2026} takes as input parameters the real-space magnetisation vector field, $M_{x,y,z}(\mathbf{r})$, calculated from the micromagnetic simulations, which are transformed into the Fourier components $\widetilde{M}_{x,y,z}(\mathbf{q})$. From these, the corresponding elastic differential spin-flip SANS cross section $d\Sigma_{\mathrm{sf}} / d\Omega$ reads~\cite{michels2021magnetic}:

\begin{widetext}
\begin{equation}
\label{eq:sf}
\frac{d\Sigma_{\mathrm{sf}}}{d\Omega} = \frac{8\pi^3}{V} b_{\mathrm{H}}^2 \left( |\widetilde{M}_x|^2 + |\widetilde{M}_y|^2 \cos^4\theta + |\widetilde{M}_z|^2 \sin^2\theta \cos^2\theta - (\widetilde{M}_y \widetilde{M}_z^* + \widetilde{M}_y^* \widetilde{M}_z) \sin\theta \cos^3\theta \right) ,
\end{equation}
\end{widetext}

where $V$ is the scattering volume, $b_{\mathrm{H}} = 2.91 \times 10^8 \, \mathrm{A}^{-1}\mathrm{m}^{-1}$ is the magnetic scattering length in the small-angle regime (the atomic magnetic form factor is approximated by $1$, since we are dealing with forward scattering), $\widetilde{\vec{M}}(\vec{q}) = \{ \widetilde{M}_x, \widetilde{M}_y, \widetilde{M}_z \}$ denotes the Fourier transform of the magnetisation vector field $\vec{M}(\vec{r}) = \{ M_x, M_y, M_z \}$, $\theta$ is the angle between $\vec{H} = H \vec{e}_z$ and $\vec{q}$, so that $\vec{q} \cong q \{ 0, \sin\theta, \cos\theta \}$ in small-angle approximation, and the asterisks ``$*$'' mark the complex-conjugated quantity. Note that the polarization-dependent chiral function has been neglected in Eq.~(\ref{eq:sf}). Since the 2D spin-flip-SANS cross section [Eq.~(\ref{eq:sf})] depends on $q_y$ and $q_z$, or equivalently on $q=(q_y^2 + q_z^2)^{1/2}$ and $\theta=\arctan(q_y/q_z)$, and can be azimuthally averaged to obtain the 1D SANS intensity $I_{\mathrm{sf}}(q)$ as:
\begin{equation}
I_{\mathrm{sf}}(q) = \frac{1}{2\pi} \int_0^{2\pi} \frac{d\Sigma_{\mathrm{sf}}}{d\Omega}(q, \theta) d\theta .
\end{equation}
To which the pair-distance distribution function \(p(r)\) is related through:
\begin{equation}
p(r) = r^2 \int_0^\infty I_{\mathrm{sf}}(q) j_0(q r) q^2 dq ,
\end{equation}
where $j_0(x) = \sin(x)/x$ denotes the zero-order spherical Bessel function. \newline

Figures~\ref{SANS}~\textbf{(B)}, \textbf{(C)}, and \textbf{(D)} show the above-mentioned spin-flip SANS observables, $\frac{d\Sigma_{\mathrm{sf}}}{d\Omega}(\theta)$, $I_{\mathrm{sf}}(q)$, and $p(r)$, for the NF texture displayed in Figure~\ref{SANS}~\textbf{(B)}. We have selected these metrics as they can reliably obtained from experimental measurements. Inspecting the P-SANS metrics, three features are jointly consistent with the presence of a vortex-like intraparticle texture: (i) a ring-like pattern in the two-dimensional spin-flip cross-section (Figure~\ref{SANS}~\textbf{(B)}); (ii) a decrease of spin-flip intensity for the low-$q$ region (Figure~\ref{SANS}~\textbf{(C)}); and (iii) a sign change in the corresponding pair-distance distribution function, $p(r)$, at a distance of approximately $r\simeq d/2=200~\mathrm{nm}$ (Figure~\ref{SANS}~\textbf{(D)}). Although these features are consistent with vortex-like configurations~\cite{adams2026angular, rai2026polarized} , none of these signatures should be interpreted as an isolated and unique proof of a vortex state. As a reciprocal-space probe, SANS is intrinsically affected by the inverse-mapping problem: a given two-dimensional spin-flip scattering cross-section does not necessarily correspond to a unique real-space magnetisation vector field. Adams \textit{et al.}~\cite{adams2026angular} showed analytically and numerically that vortex states can generate a broad \textit{zoo} of spin-flip patterns, including rings and horizontal or vertical lobes, depending on the vortex orientation, particle morphology, field history, and averaging conditions. Conversely, qualitatively similar scattering patterns may arise from distinct classes of non-uniform magnetisation. As such, the reconstruction of the underlying magnetisation therefore requires prior information concerning the particle morphology, size distribution, internal structure, magnetic interactions, and field history. \newline

What is the current state-of-the-art of experimental P-SANS? Despite substantial progress in instrumentation, theory, and numerical and experimental analysis, SANS experiments with polarization analysis still impose high technical demands. The limited neutron flux available after polarization analysis extends the counting times to hours in order to resolve weak spin-flip signals. Additional instrumental smearing, particle polydispersity, and the coexistence of intraparticle and interparticle scattering can lead to the loss of subtle texture-specific features, even when the dominant anisotropy of the scattering pattern is well resolved. The work by Rai \textit{et al.}~\cite{rai2026polarized} constitutes the most recent example at the forefront of experimental P-SANS probing non-homogeneous spin textures in NF ensembles. Although their interpretation still requires support of complementary real-space evidence of vortex configurations obtained by transmission X-ray microscopy in closely related nanoparticle systems~\cite{moya2024unveiling}, micromagnetic simulations~\cite{jefremovas2026coercivity}, and macroscopic magnetometry. Importantly, their study moves P-SANS beyond the detection of generic spin disorder~\cite{zakutna2020field} towards the experimental identification of non-homogeneous intra-particle textures, enabling the way to resolve ensemble-averaged magnetic correlations related to these textures. It thus places P-SANS among the emerging techniques capable of resolving intra-particle magnetisation textures, particularly when reciprocal-space signatures are interpreted within a multimodal framework combining real-space imaging, micromagnetic calculations, and macroscopic magnetic observables. \newline


\paragraph*{\textbf{Challenges}}
In our view, progress towards quantitative intraparticle magnetisation mapping requires advances along four interconnected directions:

\paragraph*{\textbf{Broader reciprocal-space coverage.}}
Broadening the accessible $q$-range in both low- and high-$q$ horizons would enable access to more sample information. Extending measurements towards lower $q$ is required to capture correlations associated with large multicore aggregates, collective vortex textures, and interparticle organization, whereas access to larger $q$ is needed to resolve grain-scale disorder and diffuse magnetic interfaces. For example, a minimum momentum transfer of $q_{\min}\simeq0.03~\mathrm{nm}^{-1}$ corresponds to a nominal real-space scale of approximately $2\pi/q_{\min}\simeq200~\mathrm{nm}$. This can be insufficient for fully sampling magnetic correlations in particles or aggregates several hundred nanometres in size. Very-small-angle neutron scattering, ultra-small-angle neutron scattering, and spin-echo-based SANS methods provide potential routes towards larger real-space length scales~\cite{barker2022very, schaefer2004ultra, li2021probing}. However, combining these approaches with polarization analysis, sufficient neutron flux, applied magnetic fields, and variable-temperature environments remains technically demanding. Future instrumentation should aim not only for lower $q_{\min}$, but for overlapping reciprocal-space ranges that allow intraparticle and interparticle contributions to be separated within a single, internally consistent experiment. \newline

\paragraph*{\textbf{Closing the experiment--analysis loop.}}
Polarized-SANS data reduction requires polarization-efficiency and spin-leakage corrections, detector normalization, background subtraction, absolute calibration, masking, and the selection of integration sectors and reciprocal-space regions of interest. Many of these decisions remain user dependent and are often finalized only after the beamtime has ended. This delays feedback between data quality and experimental design and can result in the most informative field, temperature, or $q$ region being undersampled. Automated and uncertainty-aware reduction pipelines could provide corrected preliminary cross-sections during the experiment, allowing acquisition parameters to be adjusted ``on-the-fly'' while the sample remains on the instrument. Machine-learning approaches may assist in identifying artifacts, weak anisotropies, or under-sampled reciprocal-space regions, but should complement rather than replace physically transparent reduction procedures. Particularly valuable would be algorithms that propagate uncertainties and flag ambiguous processing choices instead of returning a single apparently definitive scattering pattern.\newline

\paragraph*{\textbf{Standardization and interoperability.}}
Data-reduction environments such as GRASP (traditionally used at the Institut Laue-Langevin~\cite{dewhurst2023graphical}) and Mantid (usually employed at ISIS~\cite{arnold2014mantid}) include instrument-specific geometries, calibration procedures, and operational knowledge. This specialization is essential and should not simply be replaced by a single universal interface. Nevertheless, differences in metadata structures, correction procedures, output conventions, and analysis workflows complicate comparisons between experiments performed at different facilities. The priority should therefore be interoperability rather than strict software uniformity. Common metadata standards, documented correction conventions, version-controlled reduction histories, and facility-independent export formats would allow the particularities of each instrument to be retained while making reduced datasets reproducible and transferable. Coordinated European initiatives in neutron-data infrastructure, as the incoming European Spallation Source, provide an opportunity to establish these standards before increasingly large and complex polarized-scattering datasets become fragmented across incompatible analysis environments.\newline

\paragraph*{\textbf{Reference libraries and uncertainty-aware inversion.}}
A central requirement for interpreting polarized SANS is the systematic connection between classes of real-space magnetisation textures and their reciprocal-space signatures. Micromagnetic simulations can generate reference libraries spanning particle morphology, grain size, anisotropy distributions, exchange coupling, vortex orientation, field history, and interparticle organization. These libraries could support physics-informed machine-learning approaches that rank the magnetic configurations compatible with a measured dataset. The objective should not be to infer a single visually plausible texture, but to identify the family of magnetisation states consistent with the scattering observables and to quantify which additional measurement would most efficiently discriminate between them. Such an approach would transform micromagnetic modeling into an active component of experimental design.\newline

Finally, advances in three-dimensional magnetic imaging may provide complementary real-space constraints. X-ray vector nanotomography combines the penetration depth of hard X-rays with tomographic reconstruction of the three magnetisation components, enabling three-dimensional magnetic imaging of nanoscale ferromagnetic structures~\cite{donnelly2017three, chiliquinga2026three, karpov2024high}. Its application to isolated MNPs remains challenging because the dimensions and magnetic volume of the particles approach the spatial-resolution and signal-to-noise limits of the technique, which can reach approximately $10~\mathrm{nm}$ under favourable conditions~\cite{de2021fast}. Future improvements in coherent photon flux, X-ray optics, reconstruction algorithms, and correlative sample preparation may bring individual complex MNPs within reach. The most transformative outcome would not be the replacement of polarized SANS by three-dimensional imaging, but their combination: vector nanotomography could constrain the magnetisation of selected individual particles, while polarized SANS would determine whether those configurations are statistically representative of macroscopic ensembles under applied fields and variable temperatures. Such cross-validation would provide a route from real-space images and reciprocal-space correlations towards experimentally constrained, three-dimensional models of intraparticle magnetisation. \newline


\section{From forward modelling to inverse design of disorder}
The view of disorder as a potentially useful degree of freedom instead of an unavoidable deviation from an ideal crystal\cite{lak2021embracing} also changes the role of computational modelling. Until now, most advances have focused on increasingly accurate forward simulations in which particle geometry, composition and magnetic parameters are specified and the resulting behaviour is calculated. Forward modelling remains essential, but treating disorder as a functional property introduces a complementary inverse problem. Observables such as hysteresis loops, coercivity, remanence, heating efficiency, or 2D spin-flip SANS cross-sections provide incomplete and generally non-unique information about the underlying magnetic structure. This ambiguity already exists in idealized models because different combinations of anisotropy, magnetic volume, orientation, thermal activation and interactions can produce similar responses. However, disorder further increases the dimensionality and ill-posedness of the problem manyfold, as distinct defect populations and magnetic textures generate nearly indistinguishable experimental signatures.\newline

The objective should therefore not be to identify a single effective parameter set or internal configuration that reproduces one experiment. Instead, inverse modelling should determine which classes of structures remain compatible with all available observations, quantify the uncertainty and correlations within those solutions, and identify the measurements that would best distinguish between competing models. MNP modelling would thus expand from forward prediction towards forward-model-constrained inference and design, reflecting a broader shift towards inverse problems and uncertainty-aware discovery across scientific disciplines\cite{Wang2023Nature}.\newline

Recent developments in micromagnetic software provide important technical foundations for this transition. The PyTorch-based magnum.np framework integrates automatic differentiation with micromagnetic simulations\cite{Bruckner2023MagnumNP}, while NeuralMag was developed explicitly for inverse micromagnetics and adjoint-state optimization\cite{Abert2025NeuralMag}. These approaches allow gradients of an objective function to be propagated through a simulation, enabling efficient optimization of selected material parameters, magnetic states and design variables.\newline

The potential of inverse micromagnetic optimization has already been demonstrated in magnonics where it was used to design demultiplexers, nonlinear switches and circulators by specifying the desired response and allowing the algorithm to identify a suitable magnetic geometry\cite{Wang2022InverseMagnonics}. More recently, Voronov et al. combined NeuralMag with level-set topology optimization\cite{Voronov2025Topology}. In a proof-of-concept example, they modified the shape of a Stoner–Wohlfarth particle until its simulated magnetisation curve approached a prescribed target, thereby demonstrating that micromagnetic simulations can be embedded directly in gradient-based optimization loops. This is an important step because hysteresis and switching processes involve abrupt transitions that generate highly non-smooth optimization landscapes.\newline

Such approaches open the possibility of using computation as a form of ``computational microscope'', providing access to nanoparticle properties that are otherwise experimentally inaccessible \cite{Krenn2022}. A first objective is therefore the reconstruction of internal magnetic disorder from complementary measurements. Magnetic characterization probes static and dynamic responses, while techniques such as SANS are sensitive to nanoscale spin correlations and magnetic inhomogeneities in particle ensembles. Together with physics-informed optimization and machine learning, these data could enable probabilistic reconstructions of the internal magnetic states compatible with the observations. As a proof of concept, Suess et al. recently demonstrated the reconstruction of magnetic textures and global material parameters from synthetic stray-field data using machine-learning-assisted micromagnetic optimization \cite{Suess2025Reconstruction}.\newline

Once reliable structure-property relationships have been established, a more ambitious opportunity emerges: the inverse design of disorder itself. Rather than interrogating which magnetic behaviour follows from a given structure, one could ask which grain structure, defect density, oxidation profile, or disorder landscape yields a desired functional response. Extending optimization from geometry to spatially varying magnetic properties is mathematically feasible but physically far more demanding. Parameters such as saturation magnetisation, exchange, and anisotropy are coupled through composition, crystal structure, strain, interfaces, and defect chemistry, and therefore cannot generally vary independently within a real particle. Future inverse-design frameworks should therefore operate on physically meaningful and experimentally controllable variables, ensuring that optimized solutions correspond to realizable materials rather than purely mathematical constructs.\newline

The objective function must also reflect the application context. Magnetic heating depends on field amplitude and frequency, temperature, orientation, aggregation, immobilization, particle concentration, interparticle interactions and the relative contributions of N\'eel and Brownian dynamics\cite{RUTA2023185}. These dependencies are particularly important in magnetic particle hyperthermia\cite{perigo2015fundamentals}, where magnetic nanoparticles dissipate energy under an alternating magnetic field to generate localized heating for therapeutic applications such as, but not limited to\cite{gavilan2025magnetic}, cancer treatment\cite{mahmoudi2018magnetic}. Similar considerations arise in magnetic particle imaging (MPI)\cite{gleich2005tomographic}, a state-of-the art imaging modality that directly detects the nonlinear magnetic response of superparamagnetic nanoparticles. In MPI, particle properties influence signal strength, spatial resolution, and image fidelity indirectly through their magnetization dynamics\cite{velazquez2025}.\newline

A structure optimized for one field excitation may perform poorly under another or require fields unsuitable for biomedical use. Inverse design should therefore use robust, multi-objective functions that evaluate heating over a defined operating window while penalizing excessive field requirements, poor reproducibility, aggregation sensitivity or loss of performance after immobilization. Uncertainty in material parameters and environmental conditions should be propagated through the optimization rather than assessed only after an optimum has been found.\newline

A further step connects computational optimization to automated synthesis and characterization, following the broader logic of autonomous materials discovery adopted in the Materials Genome Initiative\cite{Tabor2018}. The design variables in such a loop should ultimately be experimentally controllable quantities, such as precursor concentrations, reaction temperature, oxidation conditions, annealing protocols or growth time. Characterization and modelling would infer the resulting disorder and connect it to application performance. The full loop would therefore link synthesis conditions to atomic and mesoscale structure, structure to effective magnetic properties, and those properties to relevant performance metrics\cite{carlton2025ranking} under application-relevant conditions.\newline

\paragraph*{\textbf{Challenges}}
Despite these promising developments, translating inverse micromagnetic methods into practical tools for nanoparticle characterization and materials design remains a challenge. The difficulty is not limited to optimization algorithms themselves, but arises from the multiscale nature of magnetic disorder, the computational cost of repeatedly evaluating realistic models, and the quality of the experimental data used to constrain them. Addressing these limitations will be essential to realize the vision outlined above.\newline

A first, computational, challenge is that spin disorder sits at the interface between several modelling scales. Atomistic approaches resolve local exchange interactions, surface effects, vacancies, and structural defects, whereas micromagnetics captures non-uniform magnetisation over larger length scales. Neither description spans the long timescales, broad particle distributions, interacting ensembles, and collective dynamics relevant to applications such as magnetic hyperthermia and magnetic particle imaging. Ensemble, kinetic and/or heat transport  models are therefore required to connect microscopic disorder to macroscopic performance. While individual micromagnetic simulations of nanoparticles are becoming increasingly tractable, atomistically resolved and thermally activated calculations remain expensive, particularly when repeated many times within inverse-design or reconstruction workflows. Multiscale atomistic-micromagnetic approaches already bridge some of these gaps, but a key question remains which microscopic information can be safely coarse-grained. Defect-rich particles are especially challenging because the local disorder responsible for enhanced dissipation may be precisely the information discarded by simplification. Developing reduced-order and surrogate models that preserve the relevant defect physics while remaining computationally efficient therefore represents a major methodological priority.\newline

A second challenge concerns the data itself. Problems with reproducibility, data stewardship and uncertainty are widespread in AI-driven materials science and nanotechnology\cite{shim2025next,Wang2023Nature}, and magnetic nanoparticle measurements are no exception: the RADIOMAG round-robin study reported uncertainties of approximately 30-40\% in hyperthermia performance metrics measured for identical samples across different laboratories\cite{wells2021challenges}, while subsequent analyses indicated that considerable variability remained even after instrumental issues had been accounted for~\cite{Hanson2026}. Similar challenges have been reported in magnetic nanoparticle imaging (MPI), where an inter-laboratory study revealed significant differences in nanoparticle quantification between instruments despite the use of standardized samples and protocols\cite{good2023inter}.

These limitations are particularly critical for inverse modelling. Physics-based priors can constrain poorly observed variables, but they cannot recover information that is fundamentally absent from the measurements. Without comprehensive metadata and uncertainty estimates, genuine physical variability may be indistinguishable from experimental artefacts or interlaboratory differences. Progress will therefore require larger and more informative datasets, standardized protocols, FAIR data practices, benchmark inverse problems, and reproducible simulation workflows. Ultimately, the success of inverse-design frameworks will depend as much on the quality of the underlying data as on advances in computation and optimization.\newline

\paragraph*{\textbf{Conclusion}}
Looking forward, we like to imagine that the central questions that concern us may no longer be only how accurately the behaviour of a nanoparticle with a known internal structure can be predicted. Instead, the new challenges will be to determine which internal structures are compatible with a set of experiments, which measurements are needed to reduce the remaining uncertainty, and which realizable structures should be created to achieve a desired function. Recent advances in differentiable micromagnetics, uncertainty-aware inference and autonomous materials discovery, alongside with methodological advances in P-SANS and 3D magnetic microscopy provide foundations for this transition, but current results remain proof-of-concept demonstrations rather than complete solutions for realistic, spin-disordered nanoparticle ensembles.
If, however, the associated challenges can be addressed, 
a long-term goal will become feasible: to transform spin disorder from an elusive hidden variable into a quantitatively constrained, physically interpretable and eventually engineerable design parameter.\newline

\section*{Appendix}
The benchmark problem and underlying GPU-performance data used in Fig.~\ref{fig:fast_computers} are available on the mumax3 website and in the accompanying github repository \cite{mumax3website}. The complete dataset also is reproduced here in Table~\ref{tab:gpu-benchmarks}.

\onecolumngrid

\begingroup
\small
\setlength{\tabcolsep}{6pt}
\setlength{\LTleft}{0pt}
\setlength{\LTright}{0pt}

\begin{longtable}{@{\extracolsep{\fill}} r l r @{}}
\caption{Measured \textsc{mumax}$^3$ throughput for the standard 
\textsc{mumax}$^3$ benchmark problem containing 4,194,304 finite-difference 
cells. Throughput is reported in millions of cells per second, and the year 
denotes the GPU's initial release year.}
\label{tab:gpu-benchmarks}\\

\toprule
Year & GPU & Throughput (M cells s$^{-1}$) \\
\midrule
\endfirsthead

\multicolumn{3}{c}{%
\tablename\ \thetable{} -- continued from previous page%
}\\
\toprule
Year & GPU & Throughput (M cells s$^{-1}$) \\
\midrule
\endhead

\midrule
\multicolumn{3}{r}{Continued on next page} \\
\endfoot

\bottomrule
\endlastfoot

2014 & GeForce GTX 860M & 55.3 \\
2014 & GeForce GTX 970 & 120.6 \\
2014 & GeForce GTX 980 & 132.2 \\
2015 & Quadro M5000 & 142.1 \\
2015 & Tesla M40 & 197.5 \\
2015 & GeForce GTX 980 Ti & 199.6 \\
2016 & GeForce 940MX & 26.6 \\
2016 & GeForce GTX 1050 Mobile & 72.6 \\
2016 & GeForce GTX 1060 & 128.0 \\
2016 & GeForce GTX 1060 Mobile & 128.3 \\
2016 & GeForce GTX 1070 & 170.3 \\
2016 & GeForce GTX 1080 & 189.7 \\
2016 & Quadro P5000 & 204.5 \\
2016 & Tesla P40 & 239.4 \\
2016 & TITAN X (Pascal) & 274.8 \\
2016 & Tesla P100 & 362.0 \\
2016 & Tesla P100 SXM2 & 398.8 \\
2017 & GeForce MX150 & 35.0 \\
2017 & GeForce GTX 1050 Ti Mobile & 81.9 \\
2017 & GeForce GTX 1080 Ti & 275.2 \\
2017 & TITAN Xp & 341.4 \\
2017 & TITAN V & 468.8 \\
2017 & V100-SXM2-32GB & 611.9 \\
2018 & Tesla T4 & 190.4 \\
2018 & GeForce RTX 2070 & 297.8 \\
2018 & GeForce RTX 2080 & 379.8 \\
2018 & GeForce RTX 2080 Ti & 428.4 \\
2018 & GeForce RTX 2080 Ti OC & 507.2 \\
2018 & TITAN RTX & 529.6 \\
2019 & GeForce MX250 & 34.9 \\
2019 & GeForce GTX 1650 Mobile & 97.5 \\
2019 & GeForce GTX 1660 & 164.1 \\
2019 & GeForce GTX 1660 Ti & 247.4 \\
2019 & GeForce RTX 2060 & 284.7 \\
2019 & GeForce RTX 2060 SUPER & 371.5 \\
2019 & GeForce RTX 2080 SUPER & 424.5 \\
2020 & GeForce RTX 3060 Ti & 435.3 \\
2020 & RTX A6000 (ECC On) & 558.1 \\
2020 & A40 & 574.6 \\
2020 & GeForce RTX 3080 & 687.9 \\
2020 & RTX A6000 (ECC Off) & 707.8 \\
2020 & GeForce RTX 3090 & 835.9 \\
2020 & A100 & 951.5 \\
2021 & GeForce RTX 2050 Mobile & 84.9 \\
2021 & GeForce RTX 3050 Ti & 206.3 \\
2021 & RTX A4000 (ECC On) & 323.0 \\
2021 & GeForce RTX 3080 Mobile & 400.8 \\
2021 & GeForce RTX 3070 Ti Lite Hash Rate & 547.0 \\
2022 & GeForce RTX 3080 12GB & 755.8 \\
2022 & GeForce RTX 4090 & 1228.0 \\
2022 & H100 & 1980.0 \\
2023 & GeForce RTX 4050 Mobile & 176.0 \\
2023 & GeForce RTX 4060 Ti & 299.9 \\
2023 & GeForce RTX 4070 & 498.8 \\
2024 & RTX 2000 Ada & 204.1 \\
2025 & GeForce RTX 5070 Ti & 775.9 \\
2025 & GeForce RTX 5080 & 926.2 \\
2025 & GeForce RTX 5090 & 1917.1 \\

\end{longtable}
\endgroup

\twocolumngrid
\section*{Acknowledgments}
{E.M.J. acknowledges funding from the European Union’s Horizon 2020 research and innovation program under the Marie Sk{\l}odowska-Curie Actions grant agreement 101081455 -- YIA; the Institute for Advanced Studies (IAS) of the University of Luxembourg. J. L. is supported by the Ghent University special research fund (bof/baf/1y/2024/01/005 and bof/baf/1y/2025/01/017).
This publication is based upon work from COST Action CA23132 ``Magnetic Particle Imaging for next-generation theranostics and medical research'' (NexMPI), supported by COST (European Cooperation in Science and Technology).
We thank Prof. Andreas Michels for fruitful discussions and everyone within the micromagnetic community who shared mumax3 benchmark figures for the different GPUs.}


\section*{Author roles}{Conceptualization: JL and EMJ. Data curation: JL and EMJ. Formal analysis: JL and EMJ. Funding acquisition: JL and EMJ. Investigation: JL and EMJ. Methodology: JL and EMJ. Project administration: JL and EMJ. Resources: JL and EMJ. Software Supervision: JL and EMJ. Validation: JL and EMJ. Visualization: JL and EMJ. Writing – original draft: JL and EMJ. Writing – review \& editing: JL and EMJ.}

\section*{Data availability}{The data that support the findings of this study are available upon reasonable request from the authors.}


\bibliographystyle{apsrev4-2}
\bibliography{references}

\end{document}